\documentclass[a4paper]{article}

\usepackage[margin=1in]{geometry} % full-width

\usepackage{amsmath}
\usepackage{amsthm}
\usepackage{amssymb}

\usepackage[utf8]{inputenc}
\usepackage{hyperref}
\hypersetup{
	unicode,
	pdfauthor={},
	pdftitle={Integrated photonic deep neural network with end-to-end on-chip
backpropagation training},
	pdfsubject={},
	pdfkeywords={},
	pdfproducer={LaTeX},
	pdfcreator={pdflatex}
}

\theoremstyle{plain}

\theoremstyle{definition}

\usepackage{graphicx, color, xcolor}
\graphicspath{{fig/}}

\usepackage{algorithm, algpseudocode} % use algorithm and algorithmicx for typesetting algorithms
\usepackage{mathrsfs} % for \mathscr command

\usepackage{lipsum}
\usepackage{cite}

\usepackage[left]{lineno}
\usepackage[normalem]{ulem}
\usepackage[figurename=Fig.]{caption}

\newcommand{\remove}[1]{}

\title{Integrated photonic deep neural network with end-to-end on-chip backpropagation training}
\author{Farshid Ashtiani$^{1,*}$, Mohamad Hossein Idjadi$^{1}$, Kwangwoong Kim$^{1}$}

\date{
	$^1$\textit{Nokia Bell Labs, 600 Mountain Ave, New Providence, NJ 07974, USA} \\%
    $^*$farshid.ashtiani@nokia-bell-labs.com\\[2ex]%
}

\begin{document}
\maketitle

\begin{abstract}

Integrated photonic neural networks (PNNs) have demonstrated significant potential to complement the digital electronic counterparts \cite{photonic_comp,opt_comp,PNN_prospect}. Nevertheless, robust and repeatable performance of scalable integrated PNNs is directly tied to the quality of their training. Error backpropagation (BP), which relies on nonlinear activation gradient computation, is the mainstream algorithm to train digital neural networks due to its scalability, versatility, and implementation efficiency \cite{backprop_pros}. Consequently, its adoption is highly desirable for the training of scalable PNNs. Despite such benefits and due to the lack of scalable on-chip activation gradient \cite{Training_perspective}, PNNs have mostly been trained using a digital computer to run BP, which is inadequate in addressing device variations, or through gradient-free algorithms that do not fully benefit from the versatility of BP training. Here, we report the demonstration of an integrated photonic deep neural network with end-to-end on-chip gradient-descent BP training. All linear and nonlinear computations are performed on a single photonic chip, leading to scalable and robust training despite the considerable--but typical--fabrication-induced device variations. Two nonlinear data classification tasks are demonstrated in which the chip performance matches that of the ideal digital model, both in accuracy and robustness. Integrating the advantages of BP training with PNNs allows for generalization to various PNN architectures, paving the way for scalable and reliable next-generation photonic computing systems.  

\end{abstract}

	%\tableofcontents

%%%%%%%%%%%%%%%%%%%%%%%%%%%%% INTRO %%%%%%%%%%%%%%%%%%%%%%%%%%%%%%%

\section{Introduction}

Artificial intelligence and machine learning have transformed many areas of science and technology, and this transformation is only accelerating \cite{AI1,AI2,AI3}. Artificial neural networks (ANNs) that try to mimic the human brain model, are a class of machine learning models that can learn to automatically perform specific tasks \cite{ANN} in a variety of applications such as pattern recognition \cite{pattern1,pattern2,pattern3} as well as natural language processing and language models \cite{nlp1,nlp2,nlp3,llm}, through training and inference. ANNs, in the form of multi-layer perceptrons, typically consist of multiple layers of interconnected neurons, where each neuron generates an output based on the linear weighted-sum of its inputs followed by a nonlinear activation function. Although digital electronic processors will remain the primary platform for implementing ANNs, they face energy and throughput challenges due to electrical signal transmission loss and parasitic elements as well as limited clock speed. Leveraging alternative processing modalities could support the continued growth of ANNs by addressing such challenges. Low-loss signal transmission, parallel processing via wavelength, mode, and space multiplexing, and computation by propagation make optical signal processing a promising candidate \cite{photonic_comp,opt_comp,PNN_prospect}. Recently, several PNNs have been reported that are on par with or surpass the performance of digital ANNs \cite{PDNN,hybrid_photoelec,11TOPS,forward_training,lightmatter,ultrafast_dynamic}. While a general large-scale photonic processor that can compete with the advanced digital processors such as GPUs and TPUs seems to be a distant goal, some applications can specifically benefit from PNNs. For instance, in applications where the signal of interest is in the optical domain, energy efficiency and latency of the system can be improved with PNNs that eliminate the need for front-end high-speed opto-electronic and analog-digital conversions. Object recognition in images and videos \cite{PDNN,hybrid_photoelec, 11TOPS,image1,image2,image3}, nonlinearity compensation in optical fiber links \cite{fiber_nonlin}, Fourier neural network \cite{FourierNN}, signal recovery \cite{signal_recov}, and inter-channel distortion compensation in wavelength-division multiplexing optical communication systems \cite{distortion} are among such applications. 

As with digital ANNs, PNNs must first be trained to find the optimal weights (\textit{i.e.}, operating points of photonic devices) for inference, where the accuracy of inference is directly tied to the accuracy of training. PNN training can be viewed from two main aspects: the optimization algorithm and the actual implementation method, where the choice of the former affects the latter. Gradient descent error BP training is widely used in digital ANNs. In this method, that starts from the output layer and propagates backward, the gradient of the loss function with respect to the weights of each layer is calculated and used to tune the weights and minimize the loss function through several iterations (epochs). As BP training offers efficient weight update and scalability to deep neural networks, its photonic implementation is highly desired. However, it involves the computation of the gradient of the nonlinear activation function \cite{backprop}. This requires on-chip realization of both nonlinear activation and its gradient that is particularly challenging \cite{Training_perspective}. Hence, most PNNs either rely on offline BP training on a digital computer \cite{PDNN, fiber_nonlin, hybrid1,hybrid2}, or use alternative gradient-free training algorithms such as finite difference method \cite{finite}, direct feedback alignment \cite{Direct_feedback}, associative learning \cite{associative}, and forward training \cite{forward_training,fully_forward}. Recently, a hybrid photonic BP training has been demonstrated, where the photonic chip includes the linear weights of one neural layer \cite{in_situ_backprop}. Therefore, realizing a deep neural network requires re-using the same chip and employing a digital computer for nonlinear activation function and its gradient that limit scalability. In another recent work \cite{optical_backprop}, BP training is demonstrated using free-space optics where the nonlinear activation is implemented via atomic transitions in a vapor cell. Despite impressive results, physical size and energy consumption of such bench-top setups as well as the need for special materials limit their large-scale deployment compared to integrated solutions. Two key aspects highlight the challenges of the previous integrated systems. On one hand, digital gradient-based BP training, although scalable, uses an ideal network model that does not typically include accurate photonic device models which results in sub-optimal training. Even if the device models are incorporated into the digital model, it is very hard to take the effect of process and environment variations into account. On the other hand, alternative gradient-free algorithms do not offer the versatility of gradient-based BP training. 

Here we report the demonstration of the first integrated photonic deep neural network, to our knowledge, that performs the full end-to-end on-chip gradient descent BP training. Fabricated in a standard silicon photonic foundry, the chip includes forward and backward paths with multiple layers of photonic neurons with optical linear weights and opto-electronic nonlinear activations. The forward path consists of an input layer to modulate light with the input data, a hidden layer, and an output layer. The output of the forward path is fed back to the backward path to calculate the gradient of the loss function and tune the weights of the forward path using nominally identical on-chip devices. As proof of concept, we experimentally solve two nonlinear data classification problems with inference accuracy on par with that of the equivalent ideal digital models. Moreover, our on-chip training method compensates for fabrication-induced variations, and we show its robustness and superiority over digital BP training of PNN in presence and with no prior knowledge of the device-to-device variations. PNNs can significantly boost the performance of digital ANNs for select applications; however, high-quality training is crucial to their effectiveness. This work enables robust, repeatable, and scalable on-chip BP training PNN for the next generation of hybrid deep learning systems.

%%%%%%%%%%%%%%%%%%%%%%%%%%%%% ARCHITECTURE %%%%%%%%%%%%%%%%%%%%%%%%%%%%%%%

\section{Neural network training using error backpropagation}

\begin{figure}[ht!]
\centering\includegraphics[width=\linewidth]{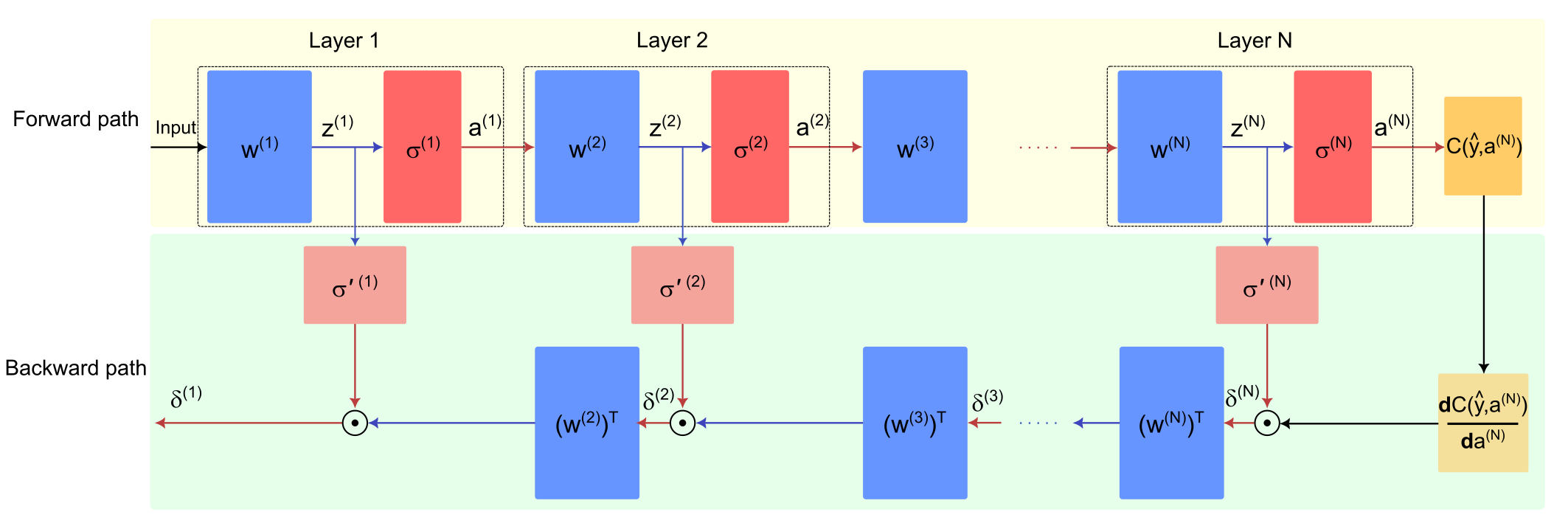}
\caption{\textbf{Backpropagation training of a neural network}. The neural network consists of $N$ layers. $w^{(l)}$, $z^{(l)}$ and $\sigma^{(l)}$ are the weight matrix, the weighted-sum output, and the nonlinear activation function, respectively. The forward path generates an output by passing the input through neural layers and the output error is calculated using the cost function $C(.)$. The backward path computes the error for each layer ($\delta^{(l)}$), starting with the output layer and backpropagating the error towards the input layer.}
\label{fig1}
\end{figure}

Figure \ref{fig1} conceptually shows the BP training process. The forward path of the neural network consists of $N$ layers, where the output of the $l^{th}$ layer ($a^{(l)}$) can be written as 

\begin{equation}
    a^{(l)} = \sigma^{(l)}(z^{(l)}) = \sigma^{(l)}(w^{(l)}a^{(l-1)}).
    \label{eq1}
\end{equation}

Here, $w^{(l)}$ is the weight matrix, $z^{(l)}$ is the weighted-sum output, and $\sigma^{(l)}$ is the nonlinear activation function of the $l^{th}$ layer. The training process starts with calculating the cost function $C(\hat{y}, a^{(N)})$, which measures the output error between the expected output $\hat{y}$ and actual output $a^{(N)}$. The main goal of gradient-descent BP training is to find the derivative of the cost function with respect to the weights of the network ($\partial{C}/\partial{w^{(l)}}$), and adjust the weights accordingly so that it minimizes the cost function. For that we define the error of the $l^{th}$ layer as \cite{BP_book}

\begin{equation}
    \delta^{(l)} = \frac{\partial{C}}{\partial{z^{(l)}}}.
    \label{eq2}
\end{equation}

In BP training, we start from calculating the error of the last layer

\begin{equation}
    \delta^{(N)} = \frac{\partial{C}}{\partial{z^{(N)}}} = \frac{\partial{C}}{\partial{a^{(N)}}}\sigma^{'(N)}(z^{(N)}).
    \label{eq3}
\end{equation}

Then, by backpropagating the error of the last layer towards the first layer, the error of the $l^{th}$ layer can be written as \cite{BP_book}

\begin{equation}
    \delta^{(l)} = ((w^{(l+1)})^{T}\delta^{(l+1)}\odot\sigma^{'(l)}.
    \label{eq4}
\end{equation}

where $T$ and $\odot$ denote matrix transpose and Hadamard (element-wise) product, respectively. Using the calculated errors, the weights of the $l^{th}$ layer can be updated using the following equation

\begin{equation}
    w^{(l)} \rightarrow w^{(l)} - \eta(\delta^{(l)}a^{(l-1)}).
    \label{eq5}
\end{equation}

where $\eta$ is the learning rate. 

\begin{figure}[ht!]
\centering\includegraphics[width=\linewidth]{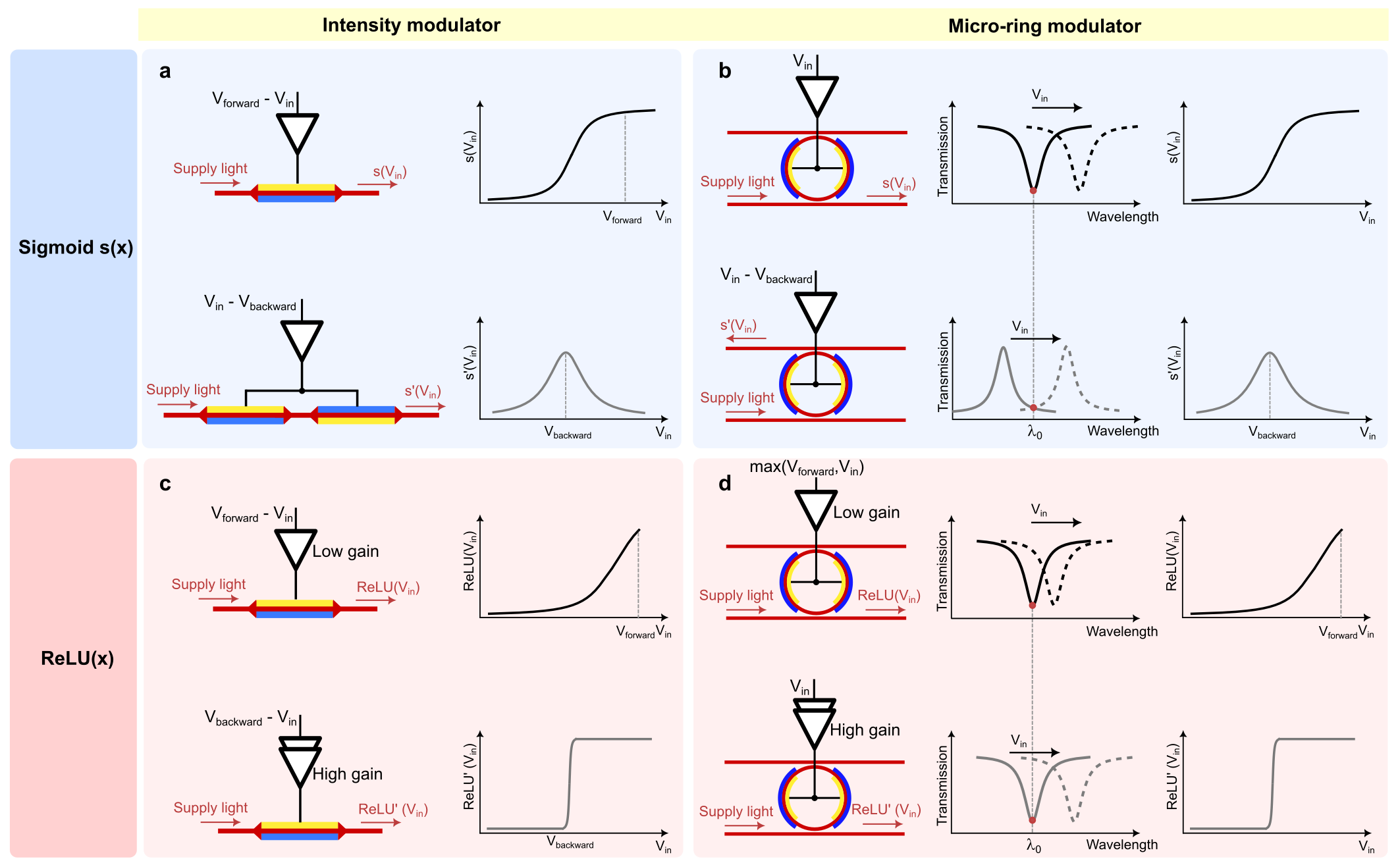}
\caption{\textbf{Photonic nonlinear activation functions and gradients using IM and MRM}. \textbf{a,} One IM is used to implement sigmoid function and two cascaded IMs approximate the gradient function. \textbf{b,} The through port of an add-drop MRM that is biased at resonance is used for sigmoid and the drop port for the gradient when the ring is biased off-resonance using $V_{backward}$. ReLU and it gradient (step function) can be implemented by switching the electrical gain using \textbf{c,} an IM initially biased at high attenuation, and \textbf{d,} a MRM that is biased at resonance.} 
\label{fig2}
\end{figure}

It can be seen that, to perform on-chip BP training, linear weight and sum, and nonlinear activation, as well as its gradient, are required. Integrated photonic implementation of the linear computation is more straightforward and has been demonstrated using a variety of approaches such as using P-doped-intrinsic-N-doped (PIN) modulators \cite{PDNN}, PN phase modulators in a Mach-Zehnder interferometer (MZI) \cite{11TOPS}, network of MZIs with thermal phase shifters \cite{forward_training}, phase change materials \cite{image1}, and micro-ring modulators \cite{fiber_nonlin} to name a few. The more challenging part is the realization of the nonlinear activation and its gradient. Despite several demonstrations of on-chip photonic activation function, to the best of our knowledge, fully integrated end-to-end photonic implementation of BP training has not been reported yet, mainly due to the need for both nonlinear activation and its gradient. In this work we propose multiple techniques for realization of these functions using standard silicon photonic components. 

Among several activation functions used in neural networks, sigmoid and rectified linear unit (ReLU) functions, and their variations, are very widely employed. Figure \ref{fig2} shows our proposed opto-electronic schemes to approximate these functions and their gradients using intensity modulators (IMs) and micro-ring modulators (MRMs). As shown in Fig. \ref{fig2}a, in the forward path, an IM can be biased at high attenuation via a bias voltage ($V_{forward}$) and as the input voltage $V_{in}$ increases, optical attenuation drops which results in a sigmoid-like function. The gradient function in the backward path can be realized using two cascaded IMs with opposite polarities (shown with different colors) simultaneously driven by the input with an offset voltage (\textit{i.e.}, $V_{in} - V_{backward}$). For input values around $V_{backward}$, attenuation by both IMs is minimum, and increasing/decreasing input results in higher attenuation. We can also use an add-drop MRM to approximate sigmoid and its gradient as shown in Fig. \ref{fig2}b. The MRM is biased on-resonance (off-resonance using $V_{backward}$) and the output taken from through (drop) port implements sigmoid (gradient) function. A similar approach applies to ReLU approximation. A single IM can be used for the ReLU and its gradient which is a step function (Fig. \ref{fig2}c). The IM is initially biased at high attenuation in both cases and the only difference is the amplifier gain that is much higher for ReLU gradient. Similarly, the MRM is biased at resonance and by switching the electrical gain the through output approximates ReLU and its gradient (Fig. \ref{fig2}d). In both cases of ReLU approximation, $V_{forward}$ should be properly adjusted to avoid the flattening of the response. It is important to mention that other variations of sigmoid function \cite{sigmoid_var}, such as logistic sigmoid, hyperbolic tangent (tanh), and arc-tangent Function (arctan), as well as ReLU variations \cite{ReLU_var} like leaky ReLU, parametric ReLU, and exponential linear unit (ELU) can be implemented using the the proposed architectures. 

Both IM and MRM are available in silicon photonic fabrication processes and can be used to implement ReLU and sigmoid functions in a PNN. In this work and as proof of concept, we utilize IM to implement ReLU activation and its gradient (Fig. \ref{fig2}c) in a photonic deep neural network and demonstrate full on-chip BP training and inference. Nevertheless, more details and experimental verification of all four implementations of Fig. \ref{fig2} are presented in Supplementary Notes.

%%%%%%%%%%%%%%%%%%%%%%%%%%%%% CHIP %%%%%%%%%%%%%%%%%%%%%%%%%%%%%%%

\section{Photonic deep neural network implementation}

\begin{figure}[ht!]
\centering\includegraphics[width=\linewidth]{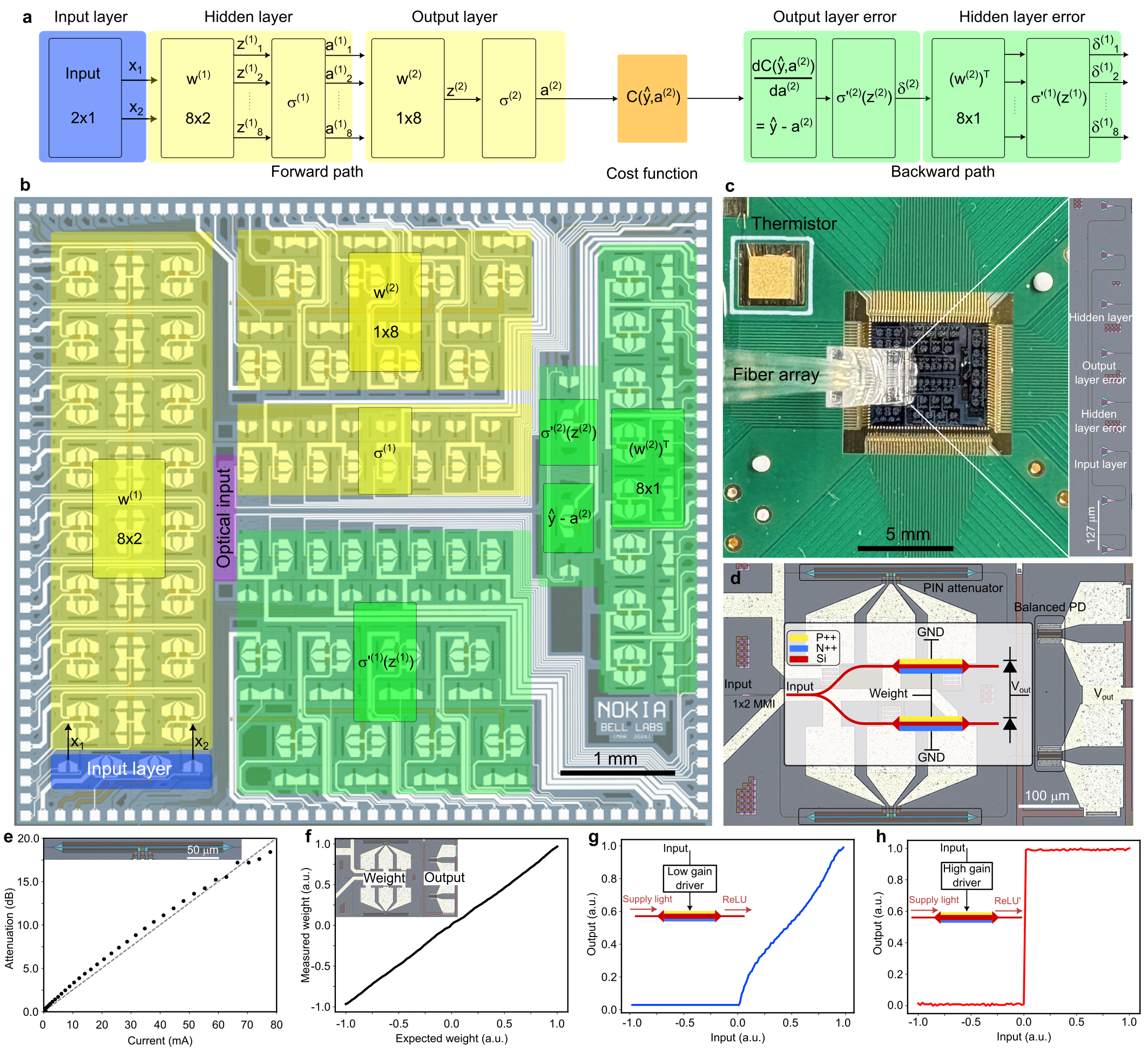}
\caption{\textbf{Silicon photonic deep neural network chip}. \textbf{a,} Schematic of the neural network architecture implemented on the silicon chip. The forward path consists of an input layer, a hidden layer and an output layer. The backward path calculates the errors corresponding to the hidden and output layers (\textit{i.e.}, $\delta^{(1)}, \delta^{(2)}$). \textbf{b,} Silicon photonic chip photograph showing different network layers. \textbf{c,} Wirebonded and packaged chip with an 8-channel fiber array and a thermistor. Inset shows the grating coupler array. \textbf{d,} Chip photo and schematic of the linear weight. Two PIN attenuators are used in a push-pull configuration to implement positive and negative weights after balanced PD. \textbf{e,} Measured PIN attenuation as a function of injected current. \textbf{f,} Measured linear relationship between the input and output of the weight unit of panel \textbf{d}. Measured response of the \textbf{g,} ReLU (with low-gain amplifier) and \textbf{h,} its gradient (with high-gain amplifier), each implemented using a single PIN attenuator.} 
\label{fig3}
\end{figure}

Figure \ref{fig3}a shows the schematic of the implemented neural network, consisting of a forward path for inference, and a backward path for error calculation and BP training. The forward path starts with the input layer where two inputs, $x_{1}$ and $x_{2}$, normalized between 0 and 1, are generated using two PIN attenuators. $x_{1}$ and $x_{2}$ are fully connected to the eight two-input weights ($w^{(1)}: 8\times2$) of the hidden layer, generating eight weighted-sum outputs ($z^{(1)}_i, i=[1,8]$). Then, $z^{(1)}_1$ to $z^{(1)}_8$ pass through eight identical IM-based ReLU-like activation functions ($\sigma^{(1)}$), resulting in the outputs of the hidden layer ($a^{(1)}_i, i=[1,8]$). The ReLU-like functions are generated using the circuit shown in Fig. \ref{fig2}c. Each output of the hidden layer is connected to a linear weight unit of the output layer ($w^{(1)}: 1\times8$), generating the weighted-sum of the eight inputs ($z^{(2)}$). The output, after passing through another ReLU activation function $\sigma^{(2)}$, is used to calculate the cost function $C(\hat{y},a^{(2)})$. Note that $\sigma^{(2)}$ and $C(\hat{y},a^{(2)})$ are implemented in the microcontroller unit. In this work, we use the mean squared error (MSE) as the cost function. Therefore, for $n$ input data points, the cost function is calculated as

\begin{equation}
    C(\hat{y},a^{(2)}) =  \frac{1}{n}\sum_{i=1}^{n}{(\hat{y}_{i} - a^{(2)}_{i})^2}  .
    \label{eq6}
\end{equation}

As detailed in equations \ref{eq1} to \ref{eq5}, in BP training, the goal is to update the network weights to minimize the cost function. To do so, we need to calculate the error associated with each layer (in this case $\delta^{(1)}$ and $\delta^{(2)}$). According to equation \ref{eq3}, the output layer error $\delta^{(2)}$ is equal to $(\hat{y}-a^{(2)})\sigma^{'(2)}(z^{(2)})$, as shown in Fig. \ref{fig3}a. Similarly, $\delta^{(1)}_{1}$ to $\delta^{(1)}_{8}$ can be calculated using equation \ref{eq4} by calculating the Hadamard product of the $(w^{(2)})^T\delta^{(2)}$ and $\sigma^{'(1)}(z^{(1)})$ (Fig. \ref{fig3}a). After error backpropagation, the results are used to update the weights of the hidden and output layers according to equation \ref{eq5} and with fixed learning rate. On-chip training algorithm is further discussed in Supplementary Notes.

Figure \ref{fig3}b shows the photograph of chip fabricated in the Advanced Micro Foundry (AMF) silicon photonic process (see Supplementary Notes for more details). All computation blocks of Fig. \ref{fig3}a are highlighted on the chip. The picture of the packaged chip is shown in Fig. \ref{fig3}c. An 8-channel standard single-mode fiber array is attached to the chip to couple the input laser via an array of grating couplers. Four grating couplers in the middle of the array are routed to different layers of the network (Fig. \ref{fig3}c inset), and the ones at the two ends are used for fiber alignment (see Supplementary Notes for details). The thermistor is mainly used for temperature monitoring. The PNN chip operates across a large optical bandwidth and does not require active temperature control (see Supplementary Notes). Figure \ref{fig3}d illustrates the linear weight unit in which the input optical signal is equally split into two signals using a 1$\times$2 multi-mode interferometer (MMI). The signals are amplitude-modulated by two PIN attenuators connected in a push-pull configuration allowing for weights between -1 and 1 after balanced photo-detection (PD). The measured PIN attenuation as a function of the injected current is shown in Fig. \ref{fig3}e. Figure \ref{fig3}f compares the expected and measured weight values. As discussed before, a single PIN attenuator implements both ReLU function and its gradient. As shown in Fig. \ref{fig3}g, in low-gain mode, a ReLU-like behavior is observed. In the high-gain mode, a step function (ReLU gradient) is measured and shown in Fig. \ref{fig3}h.

%%%%%%%%%%%%%%%%%%%%%%%%%%%%% CHIP %%%%%%%%%%%%%%%%%%%%%%%%%%%%%%%

\section{Data classification training and inference demonstration}

\begin{figure}[ht!]
\centering\includegraphics[width=\linewidth]{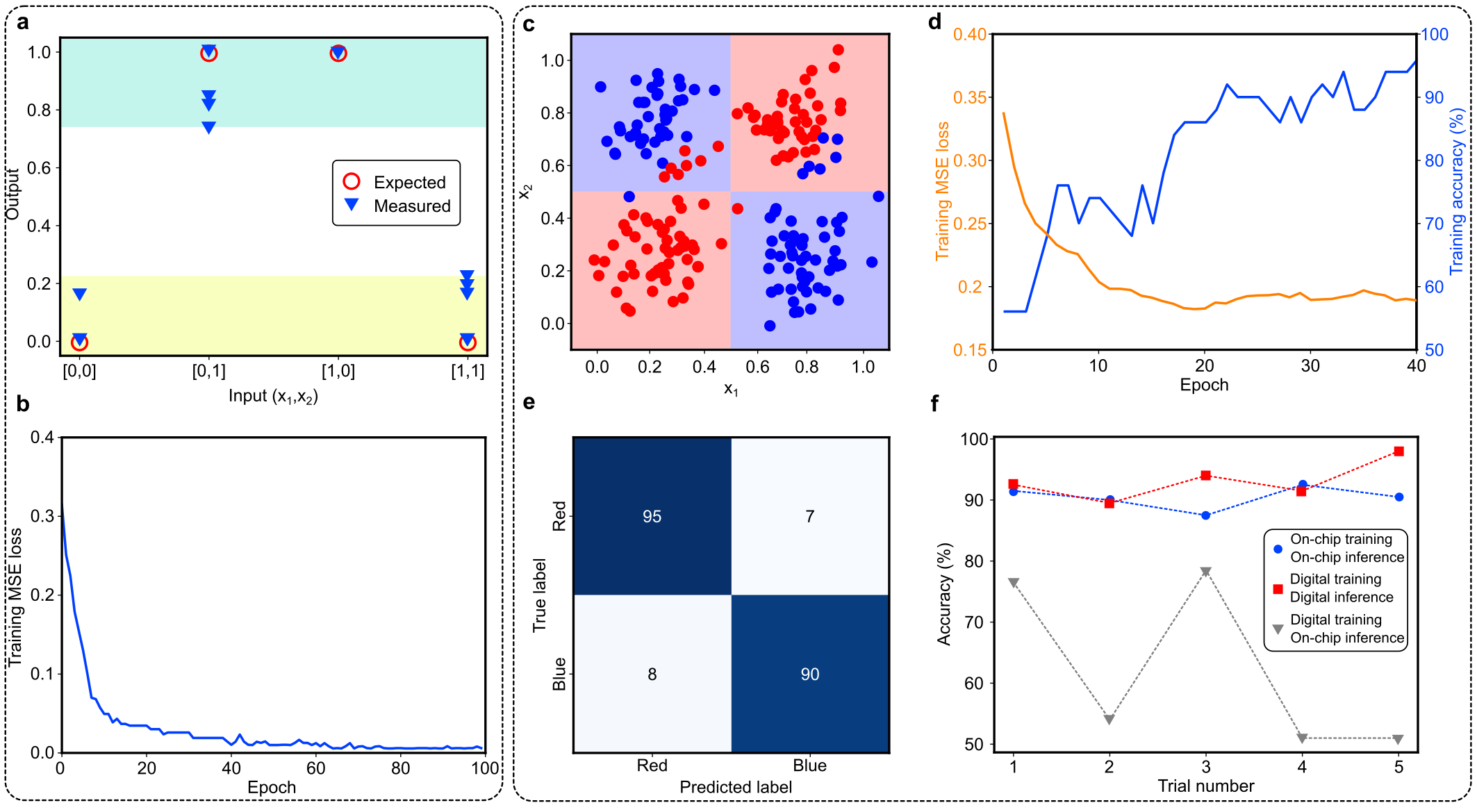}
\caption{\textbf{Nonlinear data classification demonstration}. \textbf{a,} Measured (blue triangles) and expected (red circles) XOR operation after on-chip BP training. The process is repeated multiple times and the cyan and yellow shaded regions show the variations in the output levels. \textbf{b,} MSE loss for XOR training. \textbf{c,} 2D point separation dataset. The blue and red shaded regions show true (expected) classes and points show the measured results. \textbf{d,} On-chip BP training MSE loss and accuracy and \textbf{e,} the confusion matrix corresponding to the results in panel \textbf{c}. \textbf{f,} Inference robustness for trials with different weight initializations. The graph compares end-to-end on-chip training and inference (blue), digital training and inference as the reference case (red), and digital training followed by on-chip inference (grey).} 
\label{fig4}
\end{figure}

As proof of concept, we experimentally demonstrate two nonlinear data classification problems, namely logic XOR operation and a two-dimensional (2D) point separation task. The measurement setup is explained in details in Supplementary Notes. To show the logic XOR operation, the weights are randomly initialized around zero. Once on-chip training is performed, the PNN separates the two output classes for different input pairs. This process is repeated multiple times and Fig. \ref{fig4}a shows the normalized output levels after training where clear separation between output levels can be seen. Regions corresponding to logic ``1" and ``0" are shown in cyan and yellow, respectively. Figure \ref{fig4}b plots the training MSE loss as a function of the number of epochs.

As a more complex task, we solve a 2D point separation problem. The dataset consists of 200 randomly generated $(x_{1},x_{2})$ points and the task is to separate them into blue and red points. The blue region corresponds to where $0<x_{1}<0.5$ and $0.5<x_{2}<1$ as well as where $0.5<x_{1}<1$ and $0<x_{2}<0.5$. The red region corresponds to $0<x_{1},x_{2}<0.5$ and $0.5<x_{1},x_{2}<1$. Figure \ref{fig4}c shows the blue and red shaded regions. To train the network, 50 points are randomly selected and the on-chip BP training is run for 40 epochs. Figure \ref{fig4}d shows the training loss and accuracy as a function of the number of epochs. Highest inference accuracy of 92.5\% (points shown in Fig. \ref{fig4}c) is achieved and Fig. \ref{fig4}e illustrates the corresponding confusion matrix.  

One of the major advantages of full on-chip BP training is that it can compensate for the device-to-device variations caused by fabrication and environmental effects resulting in a more robust training and inference. In contrast, it is very challenging to take such variations into account in digital training of PNNs. We discuss and show the measured on-chip device variations in Supplementary Notes. To demonstrate this robustness, we consider three scenarios. On-chip training and inference (our proposed method), digital training and on-chip inference (conventional approach), and digital training and inference (reference case). The reference case is implemented in Python using the Tensorflow library \cite{tensorflow} and the exact same neural network architecture as shown in Fig. \ref{fig3}a is implemented digitally. Figure \ref{fig4}f compares the inference accuracy of these three cases when repeated five times (trial number). In each trial, a different set of weight initializations is used. It can be seen that when the network is digitally trained (grey graph), inference accuracy varies considerably. This is expected as the training model does not incorporate actual device models and variations. Note that in a 2-class dataset, 50\% accuracy corresponds to chance-level and is considered the baseline. However, when trained using the proposed on-chip BP training, the inference is significantly more robust and repeatable, without any prior knowledge of the actual variations. Moreover, comparing the blue and red curves shows that on-chip training and inference achieves accuracy on par with the reference digital training and inference, both in terms of absolute value as well as robustness.

%%%%%%%%%%%%%%%%%%%%%%%%%%%%% DISCUSSION %%%%%%%%%%%%%%%%%%%%%%%%%%%%%%%

\section{Summary and discussion}

We demonstrated the first end-to-end on-chip photonic deep neural network BP training and inference. Linear weights, nonlinear activation, and gradient are integrated on a single silicon photonic chip. As proof of concept, we experimentally solved nonlinear inference tasks of logic XOR operation and 2D points separation with inference accuracies on par with digital models. We showed that in presence of fabrication-induced variations, the robustness and repeatability of our on-chip training addresses the variability of digital training of photonic neural networks.

The proposed architecture benefits from per-layer supply light which enables scaling to larger networks. This results in a well-defined output signal power range for each layer, independent of other layers. However, the chip in Fig. \ref{fig3} uses separate forward and backward paths that increase the number of optical inputs as each nonlinear block (main function or its gradient) requires a separate supply light. This necessitates more input optical power as well as more complex on-chip photonic routing which increases the chip area. To address this challenge, we propose a hardware reuse architecture that uses a single nonlinear block both for inference and training (the original function and its gradient), resulting in a reduction of the number of optical inputs and nonlinear blocks by a factor of two. In the modified architecture, the weight blocks also incorporate the multiplication of weight and error matrices (\textit{i.e.}, $w^{T}\delta$ in Equation \ref{eq4}) such that during inference, error values are set to one and during training the matrix multiplication is performed. The nonlinear activation blocks are mode-switched by switching the electrical connection of one IM in Fig. \ref{fig2}a, the output optical port of the MRM in Fig. \ref{fig2}b, or the gain reconfiguration in Figs. \ref{fig2}c and d, depending the type of the nonlinear function. Details of the proposed architecture are presented in Supplementary Notes. Further scalability via chip area and energy reduction can be achieved using monolithic integration of electronic and photonic components \cite{45clo,EPM_45clo}. In this case, all required electronic control circuits can be placed next to the corresponding photonic circuit. This significantly reduces parasitic components, overall system footprint, and packaging complexities. 

The focus of this work is mainly on improving training robustness and repeatability. Nevertheless, to enhance the processing speed, both for training and inference, high-speed amplitude modulation can be used. Silicon photonic MRMs enable modulation bandwidths beyond 67 GHz \cite{MRM_BW} within a small footprint, at the cost of requiring wavelength alignment and calibration. To ease the calibration requirements, compact electro-absorption modulators can be used that operate across a wide wavelength range while achieving modulation bandwidth of higher than 50 GHz \cite{EAM}. 

To fully leverage the low latency and energy efficiency of PNNs, ensuring robustness is essential. Our end-to-end on-chip BP training and inference combines the versatility of BP training algorithm, high-yield fabrication of silicon photonic chips, and performance robustness which pave the way for larger scale deployment of PNNs to boost the performance of digital processing platforms.

%%%%%%%%%%%%%%%%%%%%%%% References %%%%%%%%%%%%%%%%%%%%%%%%%

%%%%%%%%%% If using BibTeX:
%\bibliographystyle{opticajnl}
\def\url#1{}
\bibliographystyle{naturemag}
\bibliography{main}

\begin{thebibliography}{10}
\expandafter\ifx\csname url\endcsname\relax
  \def\url#1{\texttt{#1}}\fi
\expandafter\ifx\csname urlprefix\endcsname\relax\def\urlprefix{URL }\fi
\providecommand{\bibinfo}[2]{#2}
\providecommand{\eprint}[2][]{\url{#2}}

\bibitem{photonic_comp}
\bibinfo{author}{Shastri, B.~J.} \emph{et~al.}
\newblock \bibinfo{title}{Photonics for artificial intelligence and neuromorphic computing}.
\newblock \emph{\bibinfo{journal}{Nature Photonics}} \textbf{\bibinfo{volume}{15}}, \bibinfo{pages}{102--114} (\bibinfo{year}{2021}).

\bibitem{opt_comp}
\bibinfo{author}{McMahon, P.~L.}
\newblock \bibinfo{title}{The physics of optical computing}.
\newblock \emph{\bibinfo{journal}{Nature Reviews Physics}} \textbf{\bibinfo{volume}{5}}, \bibinfo{pages}{717--734} (\bibinfo{year}{2023}).

\bibitem{PNN_prospect}
\bibinfo{author}{Huang, C.} \emph{et~al.}
\newblock \bibinfo{title}{Prospects and applications of photonic neural networks}.
\newblock \emph{\bibinfo{journal}{Advances in Physics: X}} \textbf{\bibinfo{volume}{7}}, \bibinfo{pages}{1981155} (\bibinfo{year}{2022}).

\bibitem{backprop_pros}
\bibinfo{author}{Poggio, T.}, \bibinfo{author}{Banburski, A.} \& \bibinfo{author}{Liao, Q.}
\newblock \bibinfo{title}{Theoretical issues in deep networks}.
\newblock \emph{\bibinfo{journal}{Proceedings of the National Academy of Sciences}} \textbf{\bibinfo{volume}{117}}, \bibinfo{pages}{30039--30045} (\bibinfo{year}{2020}).

\bibitem{Training_perspective}
\bibinfo{author}{Buckley, S.~M.}, \bibinfo{author}{Tait, A.~N.}, \bibinfo{author}{McCaughan, A.~N.} \& \bibinfo{author}{Shastri, B.~J.}
\newblock \bibinfo{title}{Photonic online learning: a perspective}.
\newblock \emph{\bibinfo{journal}{Nanophotonics}} \textbf{\bibinfo{volume}{12}}, \bibinfo{pages}{833--845} (\bibinfo{year}{2023}).

\bibitem{AI1}
\bibinfo{author}{{National Academies of Sciences, Engineering, and Medicine}}.
\newblock \bibinfo{title}{How ai is shaping scientific discovery} (\bibinfo{year}{2023}).
\newblock \urlprefix\url{https://www.nationalacademies.org/news/2023/11/how-ai-is-shaping-scientific-discovery}.
\newblock \bibinfo{note}{Accessed: 2025-05-08}.

\bibitem{AI2}
\bibinfo{author}{Kudithipudi, D.} \emph{et~al.}
\newblock \bibinfo{title}{Neuromorphic computing at scale}.
\newblock \emph{\bibinfo{journal}{Nature}} \textbf{\bibinfo{volume}{637}}, \bibinfo{pages}{801--812} (\bibinfo{year}{2025}).

\bibitem{AI3}
\bibinfo{author}{Maslej, N.} \emph{et~al.}
\newblock \bibinfo{title}{The ai index 2024 annual report}.
\newblock \bibinfo{type}{Tech. Rep.}, \bibinfo{institution}{AI Index Steering Committee, Institute for Human-Centered AI, Stanford University}, \bibinfo{address}{Stanford, CA} (\bibinfo{year}{2024}).
\newblock \bibinfo{note}{\url{https://aiindex.stanford.edu/report/}}.

\bibitem{ANN}
\bibinfo{author}{LeCun, Y.}, \bibinfo{author}{Bengio, Y.} \& \bibinfo{author}{Hinton, G.}
\newblock \bibinfo{title}{Deep learning}.
\newblock \emph{\bibinfo{journal}{nature}} \textbf{\bibinfo{volume}{521}}, \bibinfo{pages}{436--444} (\bibinfo{year}{2015}).

\bibitem{pattern1}
\bibinfo{author}{Bishop, C.~M.}
\newblock \emph{\bibinfo{title}{Neural networks for pattern recognition}} (\bibinfo{publisher}{Oxford university press}, \bibinfo{year}{1995}).

\bibitem{pattern2}
\bibinfo{author}{LeCun, Y.}, \bibinfo{author}{Bengio, Y.} \emph{et~al.}
\newblock \bibinfo{title}{Convolutional networks for images, speech, and time series}.
\newblock \emph{\bibinfo{journal}{The handbook of brain theory and neural networks}} \textbf{\bibinfo{volume}{3361}}, \bibinfo{pages}{1995} (\bibinfo{year}{1995}).

\bibitem{pattern3}
\bibinfo{author}{Daniali, M.} \& \bibinfo{author}{Kim, E.}
\newblock \bibinfo{title}{Perception over time: Temporal dynamics for robust image understanding}.
\newblock In \emph{\bibinfo{booktitle}{Proceedings of the IEEE/CVF Conference on Computer Vision and Pattern Recognition}}, \bibinfo{pages}{5656--5665} (\bibinfo{year}{2023}).

\bibitem{nlp1}
\bibinfo{author}{Goldberg, Y.}
\newblock \bibinfo{title}{A primer on neural network models for natural language processing}.
\newblock \emph{\bibinfo{journal}{Journal of Artificial Intelligence Research}} \textbf{\bibinfo{volume}{57}}, \bibinfo{pages}{345--420} (\bibinfo{year}{2016}).

\bibitem{nlp2}
\bibinfo{author}{Goldberg, Y.}
\newblock \bibinfo{title}{A primer on neural network models for natural language processing}.
\newblock \emph{\bibinfo{journal}{Journal of Artificial Intelligence Research}} \textbf{\bibinfo{volume}{57}}, \bibinfo{pages}{345--420} (\bibinfo{year}{2016}).

\bibitem{nlp3}
\bibinfo{author}{Hirschberg, J.} \& \bibinfo{author}{Manning, C.~D.}
\newblock \bibinfo{title}{Advances in natural language processing}.
\newblock \emph{\bibinfo{journal}{Science}} \textbf{\bibinfo{volume}{349}}, \bibinfo{pages}{261--266} (\bibinfo{year}{2015}).

\bibitem{llm}
\bibinfo{author}{Arisoy, E.}, \bibinfo{author}{Sainath, T.~N.}, \bibinfo{author}{Kingsbury, B.} \& \bibinfo{author}{Ramabhadran, B.}
\newblock \bibinfo{title}{Deep neural network language models}.
\newblock In \emph{\bibinfo{booktitle}{Proceedings of the NAACL-HLT 2012 Workshop: Will We Ever Really Replace the N-gram Model? On the Future of Language Modeling for HLT}}, \bibinfo{pages}{20--28} (\bibinfo{year}{2012}).

\bibitem{PDNN}
\bibinfo{author}{Ashtiani, F.}, \bibinfo{author}{Geers, A.~J.} \& \bibinfo{author}{Aflatouni, F.}
\newblock \bibinfo{title}{An on-chip photonic deep neural network for image classification}.
\newblock \emph{\bibinfo{journal}{Nature}} \textbf{\bibinfo{volume}{606}}, \bibinfo{pages}{501--506} (\bibinfo{year}{2022}).

\bibitem{hybrid_photoelec}
\bibinfo{author}{Chen, Y.} \emph{et~al.}
\newblock \bibinfo{title}{All-analog photoelectronic chip for high-speed vision tasks}.
\newblock \emph{\bibinfo{journal}{Nature}} \textbf{\bibinfo{volume}{623}}, \bibinfo{pages}{48--57} (\bibinfo{year}{2023}).

\bibitem{11TOPS}
\bibinfo{author}{Xu, X.} \emph{et~al.}
\newblock \bibinfo{title}{11 tops photonic convolutional accelerator for optical neural networks}.
\newblock \emph{\bibinfo{journal}{Nature}} \textbf{\bibinfo{volume}{589}}, \bibinfo{pages}{44--51} (\bibinfo{year}{2021}).

\bibitem{forward_training}
\bibinfo{author}{Bandyopadhyay, S.} \emph{et~al.}
\newblock \bibinfo{title}{Single-chip photonic deep neural network with forward-only training}.
\newblock \emph{\bibinfo{journal}{Nature Photonics}} \textbf{\bibinfo{volume}{18}}, \bibinfo{pages}{1335--1343} (\bibinfo{year}{2024}).

\bibitem{lightmatter}
\bibinfo{author}{Ahmed, S.~R.} \emph{et~al.}
\newblock \bibinfo{title}{Universal photonic artificial intelligence acceleration}.
\newblock \emph{\bibinfo{journal}{Nature}} \textbf{\bibinfo{volume}{640}}, \bibinfo{pages}{368--374} (\bibinfo{year}{2025}).

\bibitem{ultrafast_dynamic}
\bibinfo{author}{Zhou, T.}, \bibinfo{author}{Wu, W.}, \bibinfo{author}{Zhang, J.}, \bibinfo{author}{Yu, S.} \& \bibinfo{author}{Fang, L.}
\newblock \bibinfo{title}{Ultrafast dynamic machine vision with spatiotemporal photonic computing}.
\newblock \emph{\bibinfo{journal}{Science Advances}} \textbf{\bibinfo{volume}{9}}, \bibinfo{pages}{eadg4391} (\bibinfo{year}{2023}).

\bibitem{image1}
\bibinfo{author}{Feldmann, J.} \emph{et~al.}
\newblock \bibinfo{title}{Parallel convolutional processing using an integrated photonic tensor core}.
\newblock \emph{\bibinfo{journal}{Nature}} \textbf{\bibinfo{volume}{589}}, \bibinfo{pages}{52--58} (\bibinfo{year}{2021}).

\bibitem{image2}
\bibinfo{author}{Wang, T.} \emph{et~al.}
\newblock \bibinfo{title}{Image sensing with multilayer nonlinear optical neural networks}.
\newblock \emph{\bibinfo{journal}{Nature Photonics}} \textbf{\bibinfo{volume}{17}}, \bibinfo{pages}{408--415} (\bibinfo{year}{2023}).

\bibitem{image3}
\bibinfo{author}{Mennel, L.} \emph{et~al.}
\newblock \bibinfo{title}{Ultrafast machine vision with 2d material neural network image sensors}.
\newblock \emph{\bibinfo{journal}{Nature}} \textbf{\bibinfo{volume}{579}}, \bibinfo{pages}{62--66} (\bibinfo{year}{2020}).

\bibitem{fiber_nonlin}
\bibinfo{author}{Huang, C.} \emph{et~al.}
\newblock \bibinfo{title}{A silicon photonic--electronic neural network for fibre nonlinearity compensation}.
\newblock \emph{\bibinfo{journal}{Nature Electronics}} \textbf{\bibinfo{volume}{4}}, \bibinfo{pages}{837--844} (\bibinfo{year}{2021}).

\bibitem{FourierNN}
\bibinfo{author}{Miscuglio, M.} \emph{et~al.}
\newblock \bibinfo{title}{Massively parallel amplitude-only fourier neural network}.
\newblock \emph{\bibinfo{journal}{Optica}} \textbf{\bibinfo{volume}{7}}, \bibinfo{pages}{1812--1819} (\bibinfo{year}{2020}).

\bibitem{signal_recov}
\bibinfo{author}{Argyris, A.}, \bibinfo{author}{Bueno, J.} \& \bibinfo{author}{Fischer, I.}
\newblock \bibinfo{title}{Photonic machine learning implementation for signal recovery in optical communications}.
\newblock \emph{\bibinfo{journal}{Scientific reports}} \textbf{\bibinfo{volume}{8}}, \bibinfo{pages}{8487} (\bibinfo{year}{2018}).

\bibitem{distortion}
\bibinfo{author}{Wang, B.}, \bibinfo{author}{De~Lima, T.~F.}, \bibinfo{author}{Shastri, B.~J.}, \bibinfo{author}{Prucnal, P.~R.} \& \bibinfo{author}{Huang, C.}
\newblock \bibinfo{title}{Multi-wavelength photonic neuromorphic computing for intra and inter-channel distortion compensations in wdm optical communication systems}.
\newblock \emph{\bibinfo{journal}{IEEE Journal of Selected Topics in Quantum Electronics}} \textbf{\bibinfo{volume}{29}}, \bibinfo{pages}{1--12} (\bibinfo{year}{2022}).

\bibitem{backprop}
\bibinfo{title}{Gradient-based learning applied to document recognition}.
\newblock \emph{\bibinfo{journal}{Proceedings of the IEEE}} \textbf{\bibinfo{volume}{86}}, \bibinfo{pages}{2278--2324} (\bibinfo{year}{1998}).

\bibitem{hybrid1}
\bibinfo{author}{Wright, L.~G.} \emph{et~al.}
\newblock \bibinfo{title}{Deep physical neural networks trained with backpropagation}.
\newblock \emph{\bibinfo{journal}{Nature}} \textbf{\bibinfo{volume}{601}}, \bibinfo{pages}{549--555} (\bibinfo{year}{2022}).

\bibitem{hybrid2}
\bibinfo{author}{Spall, J.}, \bibinfo{author}{Guo, X.} \& \bibinfo{author}{Lvovsky, A.~I.}
\newblock \bibinfo{title}{Hybrid training of optical neural networks}.
\newblock \emph{\bibinfo{journal}{Optica}} \textbf{\bibinfo{volume}{9}}, \bibinfo{pages}{803--811} (\bibinfo{year}{2022}).

\bibitem{finite}
\bibinfo{author}{Shen, Y.} \emph{et~al.}
\newblock \bibinfo{title}{Deep learning with coherent nanophotonic circuits}.
\newblock \emph{\bibinfo{journal}{Nature photonics}} \textbf{\bibinfo{volume}{11}}, \bibinfo{pages}{441--446} (\bibinfo{year}{2017}).

\bibitem{Direct_feedback}
\bibinfo{author}{Filipovich, M.~J.} \emph{et~al.}
\newblock \bibinfo{title}{Silicon photonic architecture for training deep neural networks with direct feedback alignment}.
\newblock \emph{\bibinfo{journal}{Optica}} \textbf{\bibinfo{volume}{9}}, \bibinfo{pages}{1323--1332} (\bibinfo{year}{2022}).

\bibitem{associative}
\bibinfo{author}{Tan, J.~Y.} \emph{et~al.}
\newblock \bibinfo{title}{Monadic pavlovian associative learning in a backpropagation-free photonic network}.
\newblock \emph{\bibinfo{journal}{Optica}} \textbf{\bibinfo{volume}{9}}, \bibinfo{pages}{792--802} (\bibinfo{year}{2022}).

\bibitem{fully_forward}
\bibinfo{author}{Xue, Z.} \emph{et~al.}
\newblock \bibinfo{title}{Fully forward mode training for optical neural networks}.
\newblock \emph{\bibinfo{journal}{Nature}} \textbf{\bibinfo{volume}{632}}, \bibinfo{pages}{280--286} (\bibinfo{year}{2024}).

\bibitem{in_situ_backprop}
\bibinfo{author}{Pai, S.} \emph{et~al.}
\newblock \bibinfo{title}{Experimentally realized in situ backpropagation for deep learning in photonic neural networks}.
\newblock \emph{\bibinfo{journal}{Science}} \textbf{\bibinfo{volume}{380}}, \bibinfo{pages}{398--404} (\bibinfo{year}{2023}).

\bibitem{optical_backprop}
\bibinfo{author}{Spall, J.}, \bibinfo{author}{Guo, X.} \& \bibinfo{author}{Lvovsky, A.~I.}
\newblock \bibinfo{title}{Training neural networks with end-to-end optical backpropagation}.
\newblock \emph{\bibinfo{journal}{Advanced Photonics}} \textbf{\bibinfo{volume}{7}}, \bibinfo{pages}{016004--016004} (\bibinfo{year}{2025}).

\bibitem{BP_book}
\bibinfo{author}{Nielsen, M.~A.}
\newblock \emph{\bibinfo{title}{Neural networks and deep learning}}, vol.~\bibinfo{volume}{25} (\bibinfo{publisher}{Determination press San Francisco, CA, USA}, \bibinfo{year}{2015}).

\bibitem{sigmoid_var}
\bibinfo{author}{Menon, A.}, \bibinfo{author}{Mehrotra, K.}, \bibinfo{author}{Mohan, C.~K.} \& \bibinfo{author}{Ranka, S.}
\newblock \bibinfo{title}{Characterization of a class of sigmoid functions with applications to neural networks}.
\newblock \emph{\bibinfo{journal}{Neural networks}} \textbf{\bibinfo{volume}{9}}, \bibinfo{pages}{819--835} (\bibinfo{year}{1996}).

\bibitem{ReLU_var}
\bibinfo{author}{Banerjee, C.}, \bibinfo{author}{Mukherjee, T.} \& \bibinfo{author}{Pasiliao~Jr, E.}
\newblock \bibinfo{title}{An empirical study on generalizations of the relu activation function}.
\newblock In \emph{\bibinfo{booktitle}{Proceedings of the 2019 ACM Southeast Conference}}, \bibinfo{pages}{164--167} (\bibinfo{year}{2019}).

\bibitem{tensorflow}
\bibinfo{author}{Abadi, M.} \emph{et~al.}
\newblock \bibinfo{title}{{TensorFlow}: Large-scale machine learning on heterogeneous systems} (\bibinfo{year}{2015}).
\newblock \urlprefix\url{https://www.tensorflow.org/}.
\newblock \bibinfo{note}{Software available from tensorflow.org}.

\bibitem{45clo}
\bibinfo{author}{Rakowski, M.} \emph{et~al.}
\newblock \bibinfo{title}{45nm cmos-silicon photonics monolithic technology (45clo) for next-generation, low power and high speed optical interconnects}.
\newblock In \emph{\bibinfo{booktitle}{Optical Fiber Communication Conference}}, \bibinfo{pages}{T3H--3} (\bibinfo{organization}{Optica Publishing Group}, \bibinfo{year}{2020}).

\bibitem{EPM_45clo}
\bibinfo{author}{Omirzakhov, K.}, \bibinfo{author}{Hao, H.}, \bibinfo{author}{Pirmoradi, A.} \& \bibinfo{author}{Aflatouni, F.}
\newblock \bibinfo{title}{Energy efficient monolithically integrated 256 gb/s optical transmitter with autonomous wavelength stabilization in 45 nm cmos soi}.
\newblock \emph{\bibinfo{journal}{IEEE Journal of Solid-State Circuits}}  (\bibinfo{year}{2024}).

\bibitem{MRM_BW}
\bibinfo{author}{Chan, D. W.~U.} \emph{et~al.}
\newblock \bibinfo{title}{C-band 67 ghz silicon photonic microring modulator for dispersion-uncompensated 100 gbaud pam-4}.
\newblock \emph{\bibinfo{journal}{Optics Letters}} \textbf{\bibinfo{volume}{47}}, \bibinfo{pages}{2935--2938} (\bibinfo{year}{2022}).

\bibitem{EAM}
\bibinfo{author}{Srinivasan, S.~A.} \emph{et~al.}
\newblock \bibinfo{title}{60gb/s waveguide-coupled o-band gesi quantum-confined stark effect electro-absorption modulator}.
\newblock In \emph{\bibinfo{booktitle}{Optical Fiber Communication Conference}}, \bibinfo{pages}{Tu1D--3} (\bibinfo{organization}{Optica Publishing Group}, \bibinfo{year}{2021}).

\bibitem{IPC}
\bibinfo{author}{Ashtiani, F.} \& \bibinfo{author}{Idjadi, M.~H.}
\newblock \bibinfo{title}{On-chip nonlinear activation and gradient functions for photonic backpropagation training and inference}.
\newblock In \emph{\bibinfo{booktitle}{2023 IEEE Photonics Conference (IPC)}}, \bibinfo{pages}{1--2} (\bibinfo{year}{2023}).

\bibitem{adam}
\bibinfo{author}{Kingma, D.~P.}
\newblock \bibinfo{title}{Adam: A method for stochastic optimization}.
\newblock \emph{\bibinfo{journal}{arXiv preprint arXiv:1412.6980}}  (\bibinfo{year}{2014}).

\bibitem{AMF_var}
\bibinfo{author}{Siew, S.~Y.} \emph{et~al.}
\newblock \bibinfo{title}{Review of silicon photonics technology and platform development}.
\newblock \emph{\bibinfo{journal}{Journal of Lightwave Technology}} \textbf{\bibinfo{volume}{39}}, \bibinfo{pages}{4374--4389} (\bibinfo{year}{2021}).

\end{thebibliography}
%\addbibresource{refs.bib}

%\addbibresource{refs.bib}

\section{Data availability}
The data supporting findings of this study is available from the corresponding author upon reasonable request.

% \section{Acknowledgments}

\section{Author contributions}
F.A. conceived the idea. F. Ashtiani and M.H. Idjadi taped-out the chip, designed the electronic circuit board, and conducted the measurements. K.W. Kim packaged the photonic chip. F. Ashtiani wrote the paper. All authors reviewed the paper.

\section{Competing interests}
The authors declare no competing interests.

\section{Additional information}
\textbf{Correspondence} and requests for materials should be addressed to Farshid Ashtiani.

\newpage

%%%%%%%%%%%%%%%%%%%%%%%%%%%%% Methods %%%%%%%%%%%%%%%%%%%%%%%%%%%%%%%

\section{Supplementary Notes}

\textbf{Supplementary note 1: Experimental verification of sigmoid and ReLU approximation using IM and MRM}\\ In this work and the implemented photonic chip, we used IM-based ReLU activation and its gradient as the nonlinearity and demonstrated training and inference for two classification tasks. Nonetheless, other proposed circuits shown in Fig. \ref{fig2} can also serve as activation. Therefore, we conducted a series of measurements to demonstrate all nonlinear activation approximations. Supplementary Figure \ref{ex_fig1} shows the normalized measured activations and the corresponding gradients for all proposed architectures. In all plots, blue dots and red curves represent experimental and theoretical graphs, respectively. Supplementary Figures \ref{ex_fig1}a and b show the results for sigmoid function and Supplementary Figs. \ref{ex_fig1}c and d illustrate the results for ReLU function. Decent agreement between measurement and theory can be seen. It should be noted that to demonstrate MRM-based sigmoid and ReLU functions, we used an add-drop micro-disk modulator fabricated in AIM Photonics silicon photonic process with a similar response as a typical MRM (Supplementary Figs. \ref{ex_fig1}b and d) \cite{IPC}. The IM-based results are based on the same PIN attenuators (IMs) used in the PNN chip (Supplementary Figs. \ref{ex_fig1}c and d).\\ \\
\textbf{Supplementary note 2: Training algorithm}\\ On-chip training process follows the commonly used gradient descent-based BP algorithm. Supplementary Figure \ref{ex_fig2} illustrates the training process. The training dataset is stored in the microcontroller memory and the input pairs are loaded to the input layer as a vector ($X= (x_{1},x_{2})$). In parallel, hidden layer weights are loaded and the weighted-sum vector $W^{(1)}X$ is read and transferred to the ReLU nonlinear activation to generate $a^{(1)}=\sigma(W^{(1)}X)$. The optical output is routed to the output layer weight block and the resulting weighted-sum $W^{(2)}a^{(1)}$ is stored in the microcontroller memory. After applying another ReLU nonlinearity and calculating the cost function, the gradient of output layer activation $\sigma'(z^{(2)})$ is calculated on the chip and is then multiplied by the derivative of the cost function $\hat{y}-a^{(2)}$, which results in output layer error $\delta^{(2)}$. Next, the hidden layer error is calculated using the result of the previous step to generate $\delta^{(1)}=(W^{(2)})^{T}\delta^{(2)}\odot\sigma'(z^{(1)})$. Finally, the hidden and output layer weights are updated according to equation \ref{eq5}. 

This process continues for the target number of epochs. The number of epochs is set such that the MSE loss stabilizes and does not become smaller. A learning rate ($\eta$) of 0.05 is used for on-chip BP training. Note that other variations of BP training such as adaptive learning rate, adding momentum to weight update, and Adam optimizer \cite{adam} can also be used in the proposed on-chip training. \\ \\
\textbf{Supplementary note 3: Chip fabrication}\\ The photonic chip was fabricated in the AMF SOI silicon photonic process. The platform features 3 $\mu$m buried oxide and all on-chip routing is done using single-mode waveguides that are 220 nm thick and 500 nm wide, with a measured loss of less than 2 dBcm$^{-1}$. Supplementary Table 1 shows the specifications of the on-chip photonic components. \\ \\
\textbf{Supplementary note 4: Experimental setup and control electronics}\\ Supplementary Figure \ref{ex_fig3}a shows the picture of the experimental setup. The output of a tunable laser (santec TSL-570) at 1550 nm and 7 dBm optical power is split into four signals using fiber splitters. Each of the four signal passes through a polarization controller and connected to one input of the 8-channel fiber array attached to the chip. As shown in Fig. \ref{fig3}c, the fiber array is glued to the chip and proper alignment is done using the four end grating couplers (two on each end). The other four grating couplers couple the light to the input waveguide of each layer of the network (shown in Fig. \ref{fig3}c). The on-board thermistor is used to monitor the temperature and a temperature variation of about $\pm2^{\circ}$C around room temperature ($22^{\circ}$C) was observed. No temperature control loop is used as the PIN attenuator works across a 100 nm bandwidth around 1550 nm and is not sensitive to such temperature variations. However, for implementations using resonance-based devices such as MRMs, temperature stabilization is necessary for reliable operation. 

To write/read the data to/from the photonic chip, a custom electronic circuit board is designed. Supplementary Figure \ref{ex_fig3}b shows the block diagram of the control electronics as well as a picture of the assembled board. The narrow and wide arrows represent single- and multi-signal connections, respectively. Four 12-bit 16-channel serial digital-to-analog converters (DACs) convert the data stored in the memory, including the weights and the inputs to the nonlinear activation functions and their gradients. The serial interface of the microcontroller and the DACs consists of data, clock, and chip-select signals. An array of 60 operational amplifiers (OPAMPs) is used to drive the on-chip PIN attenuator. Read-out circuitry consists of six 8-bit parallel analog-to-digital converters (ADCs) to read 45 signals coming out of the photonic chip. The interface of the ADC array and the microcontroller includes 8 data lines, one chip select, and three address lines. A complete list of the components used in the experimental setup are provided in Supplementary Table 2.\\ \\
\textbf{Supplementary note 5: On-chip device variations}\\ As discussed before, one of the primary goals of end-to-end on-chip BP training is to compensate for the device variations. Such variations, if not accounted for properly, can significantly degrade the repeatability, robustness, and accuracy of training and inference. While Fig. \ref{fig4}f shows the robustness of our implementation, it is important to study the actual device variations. Typical variations in thickness, width, and loss of a single-mode waveguide in AMF process are about 2.2\%, 1.2\%, and 17\%, respectively \cite{AMF_var}. Such variations result in considerable change in the effective index of refraction ($n_{eff}$) and the group index ($n_{g}$). Our MODE simulations show that the standard deviation of variations in $n_{eff}$ and $n_{g}$ are 0.0189 and 0.01, respectively,  affecting the performance of the on-chip photonic devices.

To find out the actual effect of such variations, we conducted detailed measurements of all on-chip linear and nonlinear computation. Supplementary Figure \ref{ex_fig4} shows the variations of different on-chip photonic computation blocks. In all graphs, the area shaded in blue shows the variation and the red line corresponds to the average. Note that Figs. \ref{fig3}f, g, and h show the average graphs. Supplementary Figures \ref{ex_fig4}a, b, c, and d correspond to the weight blocks of the hidden layer, output layer, output layer error calculation, and hidden layer error calculation, respectively. The photonic circuit for all weights is similar to the one shown in Fig. \ref{fig3}d. For weights between -1 and 1, a variation up to 0.273 can be seen which is significant. Supplementary Figure \ref{ex_fig4}e shows the multiply-accumulate (MAC) operation accuracy using the on-chip weight blocks where the mean and standard deviation of error is measured to be -0.0047 and 0.3662, respectively. The ReLU activation and gradient approximations exhibit variations that are illustrated in Supplementary Figs. \ref{ex_fig4}g and h. The variations in the gradient graphs is more significant mainly due to one of the PIN attenuators whose response is considerably different from others. Considering these measured variations, the on-chip training compensates for the errors as evident from the robustness test shown in Fig. \ref{fig4}f. In contrast, digital training does not perform robustly as it does not include the actual on-chip device variations.    \\ \\
\textbf{Supplementary note 6: Hardware reuse architecture for scaling up}\\ Per-layer supply light enables scaling as it prevents optical loss propagation across multiple layers. Each nonlinear block, both in the forward and backward paths, requires an interdependent supply light. Therefore, for an $N$-layer network, $2N$ supply lights are needed. This results in larger overall optical power consumption as well as more on-chip optical routing which increases the chip area. To address this challenge, we propose a hardware reuse architecture as shown in Supplementary Fig. \ref{ex_fig5}. The idea is to use a single nonlinear block in each layer both in inference and training. In this figure, the forward path consists of modified weight and nonlinear activation blocks such that the same blocks can be configured for training and inference. The modified weight block of the $i^{th}$ layer consists of two arrays of IMs, one to encode matrix $(\delta^{(i)})^T$ and one to encode matrix $w^{(i)}$. In the inference mode, the first array is transparent (transmission of one) and the input to the block is multiplied by the weight matrix. In BP training mode however, the error matrix $(\delta^{(i)})^T$ is loaded to the first array of modulators and the output of the modified weight block is equal to $(\delta^{(i)})^{T}w^{(i)}$. Note that equation \ref{eq4} can be written as

\begin{equation}
    \delta^{(i-1)} = ((\delta^{(i)})^{T}w^{(i)})^{T}\odot\sigma^{'(i-1)}.
    \label{eq7}
\end{equation}

Therefore, to compute $\delta^{(i-1)}$, the transpose of the output of modified weight block should be multiplied by the gradient of the nonlinear activation of the previous layer ($\sigma^{'(i-1)}$). This is performed in the Hadamard product block which is an array of parallel IMs whose optical input is $\sigma^{'(i-1)}$ and electrical input is $((\delta^{(i)})^{T}w^{(i)})^{T}$. 

The nonlinear activation block is also modified to a mode switching activation that generates the desired output based on the mode of operation (inference or training). Depending on the type of nonlinear function (Fig. \ref{fig2}), the activation block can be designed. For IM-based sigmoid activation, the electrical connection of one of the IMs can be switched to change between sigmoid and its gradient. The switch is open for inference and closed for training. In the MRM-based implementation, an optical switch can be used to take the output either from the through port (inference) or from the drop port (training). In the case of ReLU activation, a programmable gain amplifier can be utilized in both IM- and MRM-based architectures where low (high) gain is set for inference (training). All of the proposed mode switching activation circuits are schematically shown in Supplementary Fig. \ref{ex_fig5}. 

In addition to reducing the number of optical inputs by a factor of two, the number of nonlinear blocks is also half of the original design, further reducing the chip area and the number of components. Moreover, the proposed approach reduces the number of electronic components controlling the reused nonlinear blocks which in turn simplifies the system design.

\newpage

%%%%%%%%%%%%%%%%%%%%%%%%%%%%% Supplementary figure %%%%%%%%%%%%%%%%%%%%%%%%%%%%%%%

\renewcommand{\figurename}{Supplementary Fig.}
\renewcommand\thefigure{\arabic{figure}}
\setcounter{figure}{0}

\newpage
\begin{figure}[ht!]
\centering\includegraphics[width=\linewidth]{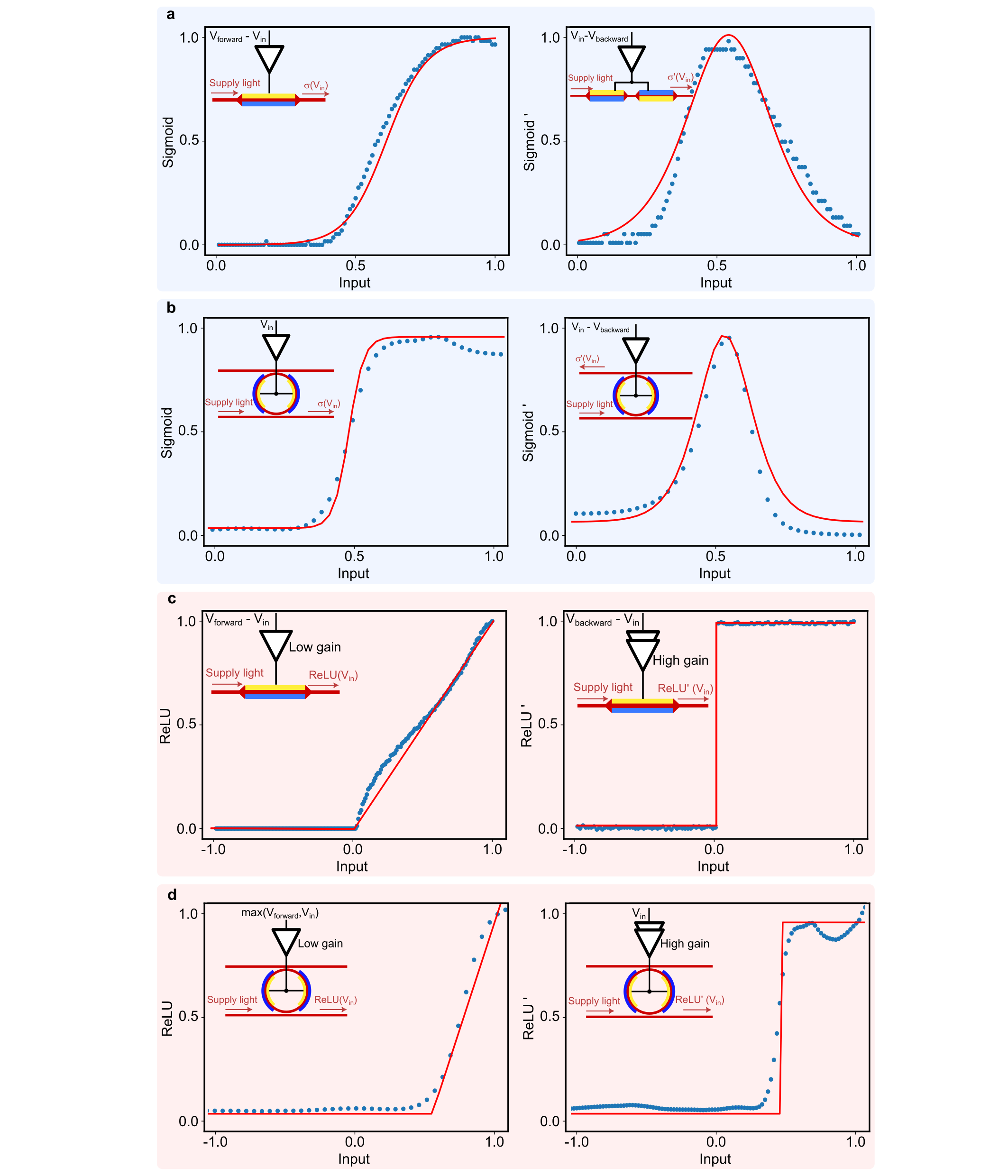}
\caption{\textbf{Photonic approximation of sigmoid and ReLU using IM and MRM.} Normalized experimental (blue dots) and theoretical (red lines) graphs of sigmoid and ReLU nonlinear activations using the architectures shown in Fig. \ref{fig2}. Sigmoid and its gradient using \textbf{a,} IM and \textbf{b,} MRM. ReLU and its gradient using \textbf{c,} IM and \textbf{d,} MRM.}   
\label{ex_fig1}
\end{figure}

\newpage

\begin{figure}[ht!]
\centering\includegraphics[width=\linewidth]{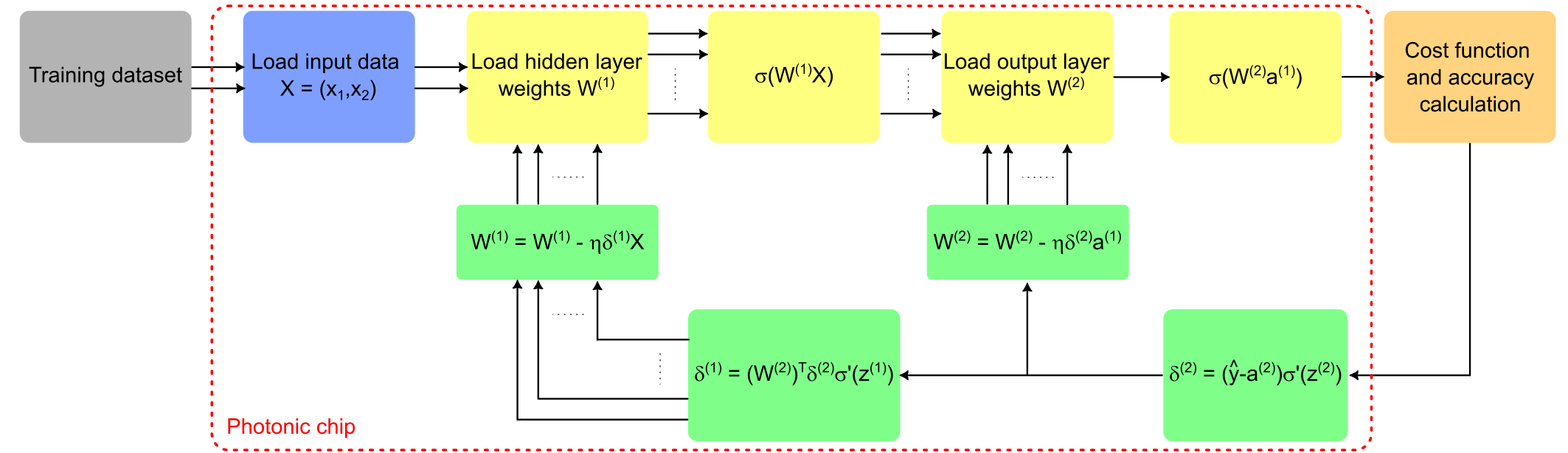}
\caption{\textbf{On-chip BP training process}. Training data and hidden layer weights are loaded their corresponding modulators. Weighted-sum output of the hidden layer ($z^{(1)} = w^{(1)}X$) is read to drive its ReLU nonlinearity and generate $a^{(1)} = \sigma(w^{(1)}X)$ that is multiplied by the output layer weight matrix. All acquired data are stored in a memory for training. Output nonlinearity and cost function are applied to the result in the microcontroller. In the backward path, weighted-sum ($z$) and neural output ($a$) signals are recalled from the memory and loaded to the corresponding modulators to calculate output and hidden layer errors ($\delta^{(1)}$ and $\delta^{(2)}$) and update the weights accordingly.}   
\label{ex_fig2}
\end{figure}

\newpage

\begin{figure}[ht!]
\centering\includegraphics[width=\linewidth]{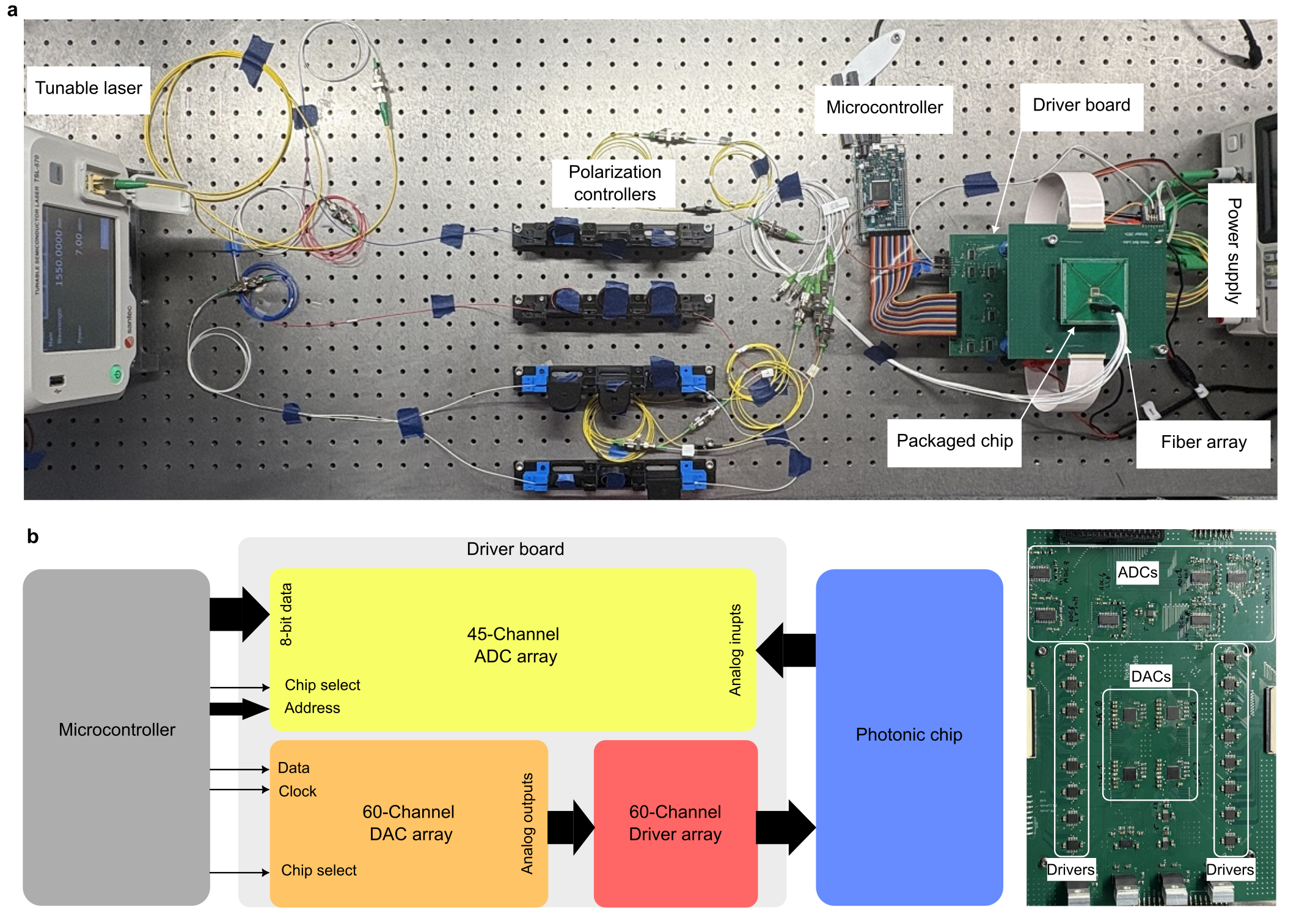}
\caption{\textbf{Experimental setup and control electronics}. \textbf{a,} Picture of the experimental setup. The output of a tunable laser is split into four signals using fiber splitters. After polarization adjustment, the four optical signals are vertically coupled to the chip using a single-mode fiber array, A microcontroller together with a custom designed electronic circuit board control the photonic chip. \textbf{b,} Block diagram and picture of the control electronic circuit consisting of an array of ADCs to read the output signals, and an array of DACs followed by drivers to write the data to the photonic chip.} 
\label{ex_fig3}
\end{figure}

\newpage

\begin{figure}[ht!]
\centering\includegraphics[width=\linewidth]{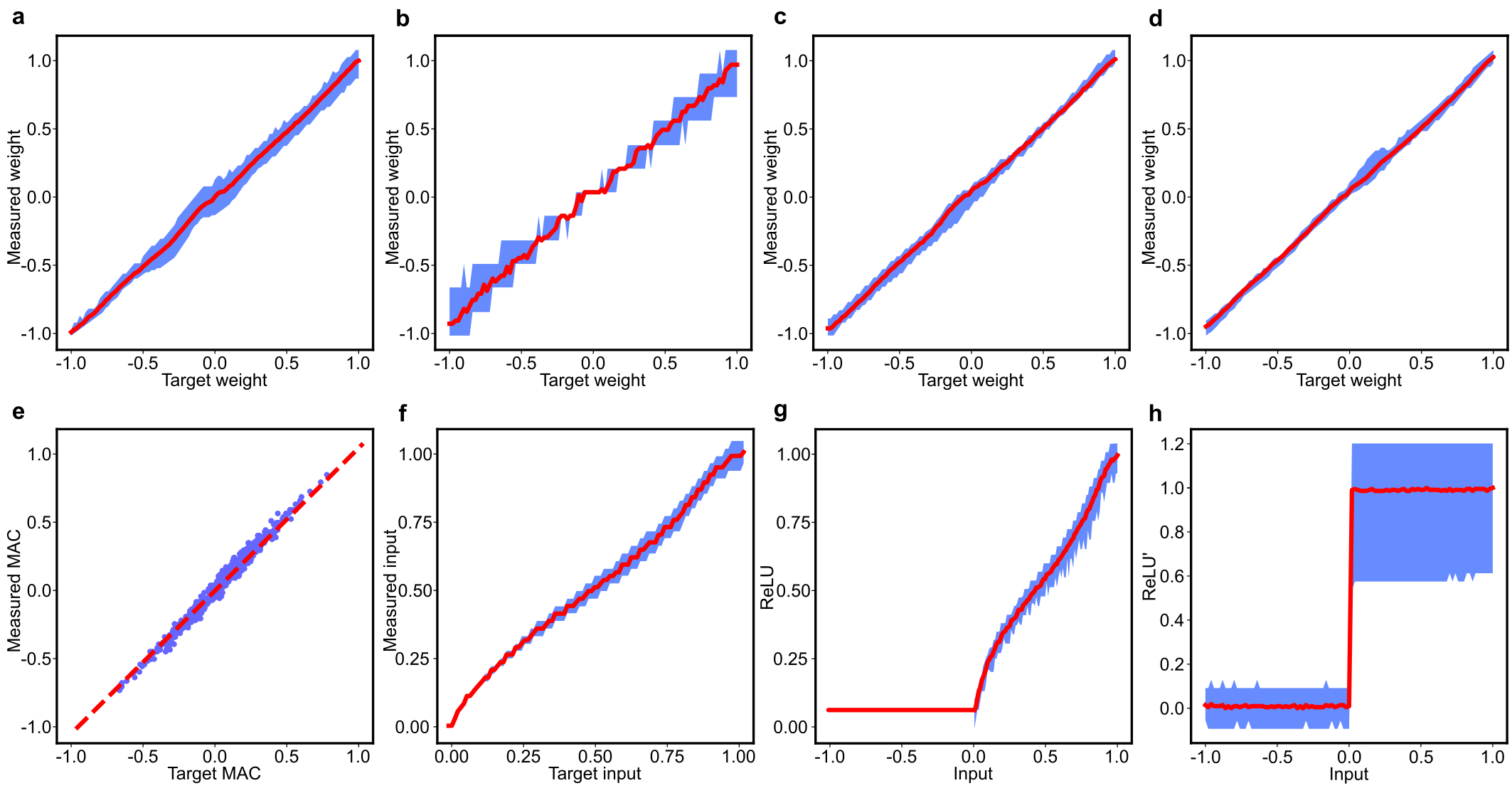}
\caption{\textbf{On-chip device variations}. Measured variations of various neural network layer blocks. \textbf{a,} Hidden layer weights. \textbf{b,} Output layer weights. \textbf{c,} Output layer error calculation weights. \textbf{d,} Hidden layer error calculation weights. \textbf{e,} Accuracy of MAC operation used in weight blocks. \textbf{f,} Input layer data mapping. \textbf{g,} ReLU nonlinear activation and \textbf{h,} its gradient.} 
\label{ex_fig4}
\end{figure}

\newpage

\begin{figure}[ht!]
\centering\includegraphics[width=\linewidth]{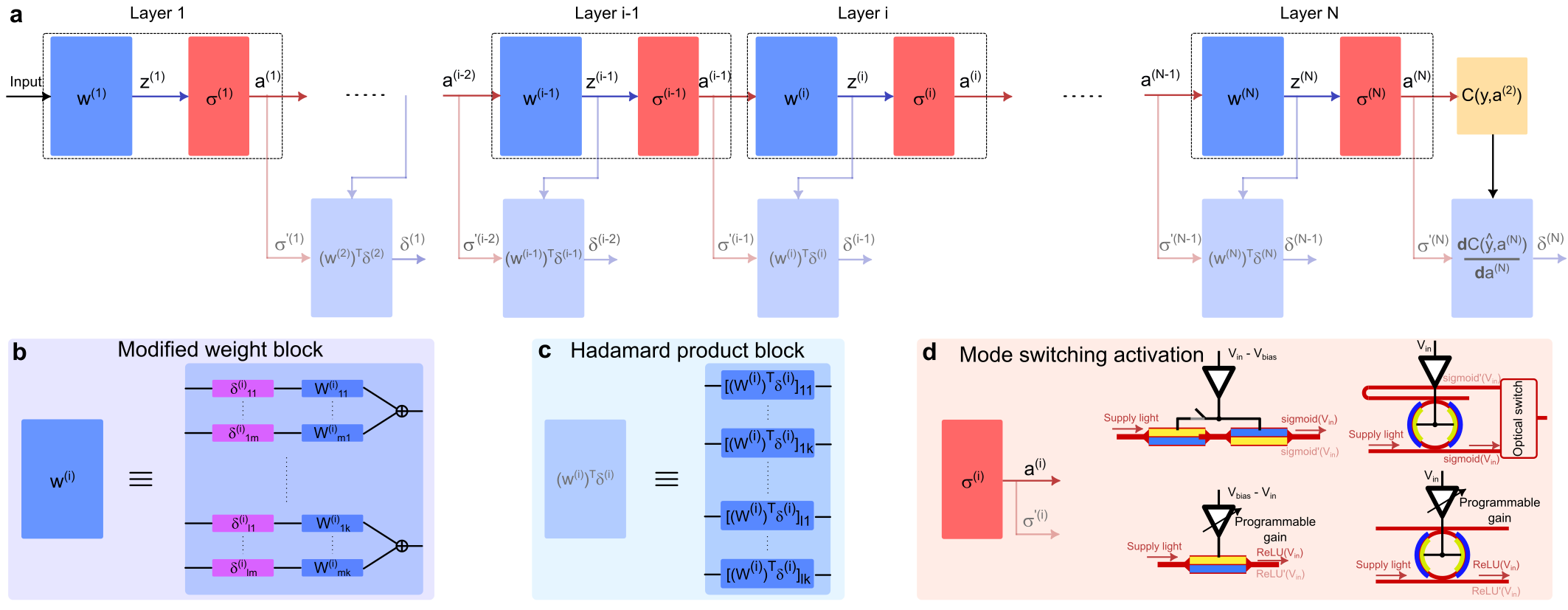}
\caption{\textbf{On-chip BP training via hardware reuse for improved scaling}. \textbf{a,} Block diagram of an N-layer neural network with N reused nonlinear blocks. \textbf{b,} Linear weight blocks are modified to include modulators for layer error values (\textit{i.e.}, $\delta^{i}$). These modulators are transparent (full transmission mode) during inference and are used in training to calculate $(\delta^{i})^{T}w^{i}$. \textbf{c,} Hadamard product block consisting of an array of weights similar to that of Fig. \ref{fig3}d to calculate $((w^{(i)})^{T}\delta^{(i)}\odot\sigma^{'(i)}$. \textbf{d,} Mode switching nonlinear activation. Depending on the type of nonlinear activation, a single block operates in training and inference modes by switching electrical connection, gain, and output port.} 
\label{ex_fig5}
\end{figure}

\newpage
Supplementary Table 1: \textbf{On-chip device specifications} 
\begin{figure}[ht!]
\centering\includegraphics[width=\linewidth]{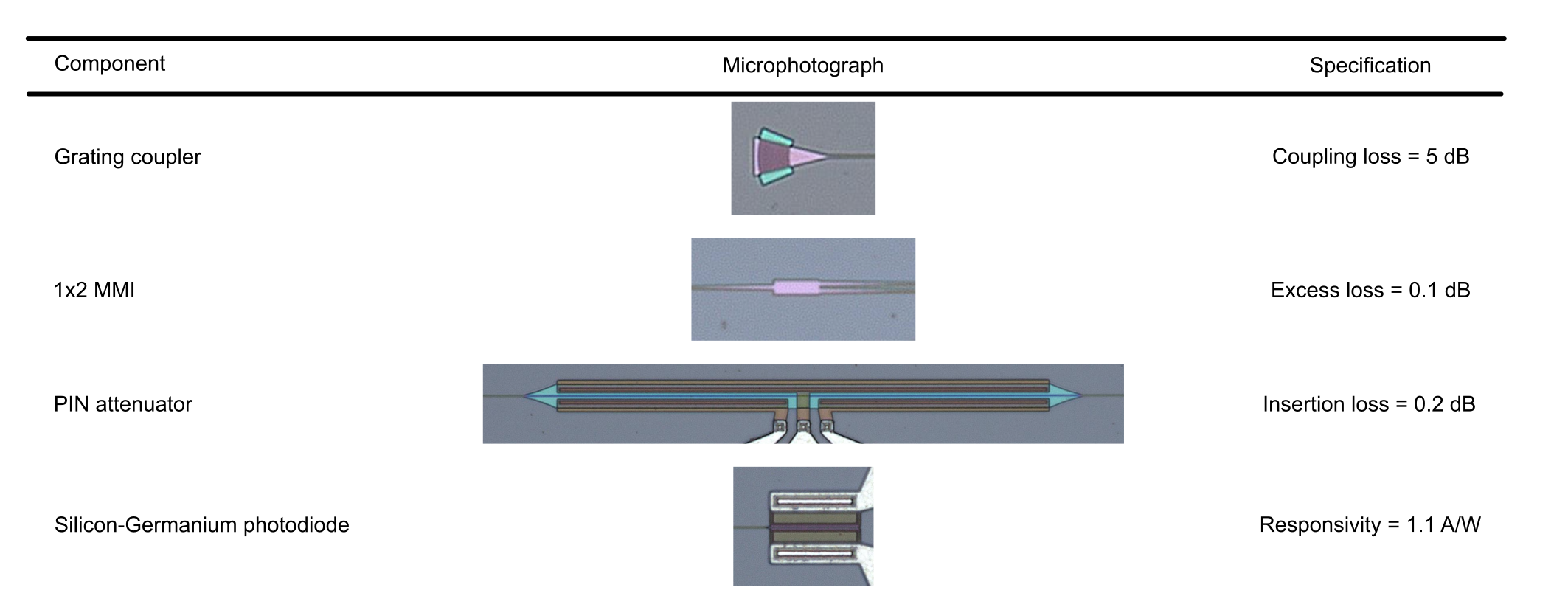}
\label{ex_tb1}
\end{figure}

\newpage
Supplementary Table 2: \textbf{Experimental setup equipment list}
\begin{figure}[ht!]
\centering\includegraphics[width=\linewidth]{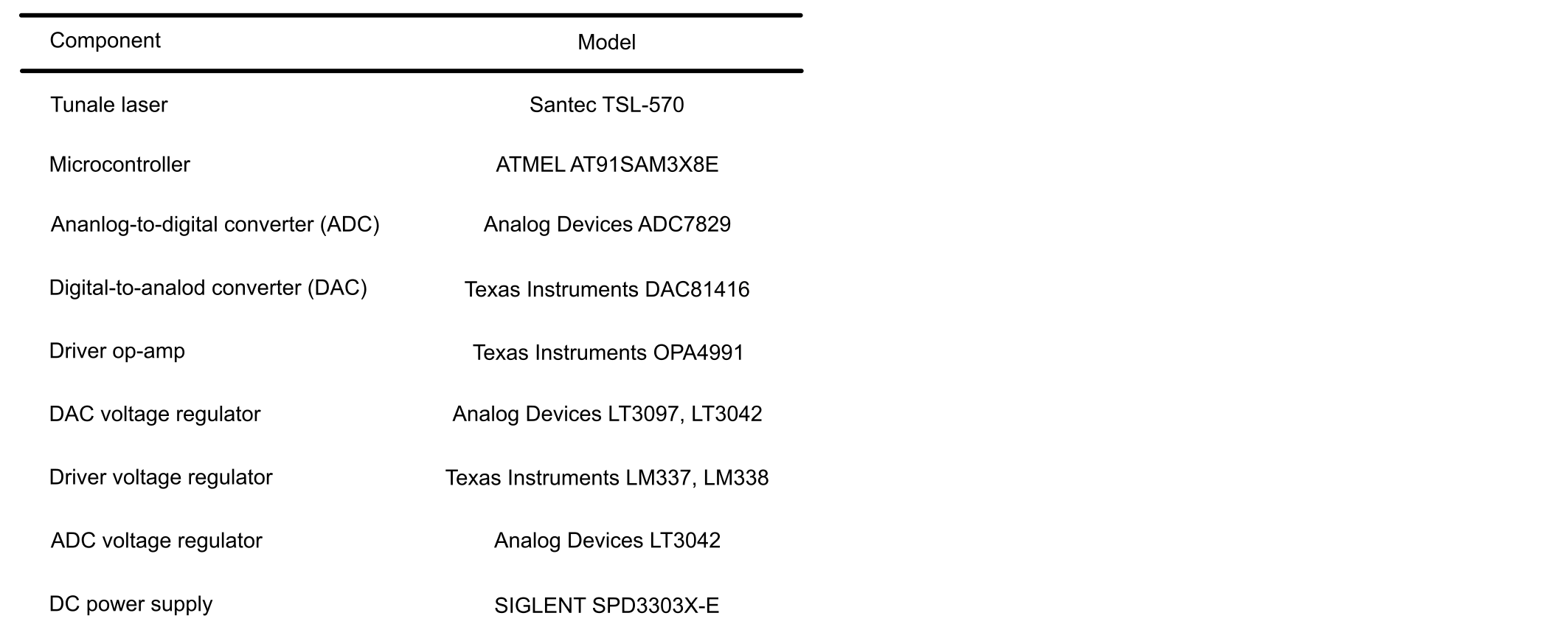}
\label{ex_tb2}
\end{figure}

\end{document}